# Functional dual-slope frequency-domain near-infrared spectroscopy data interpreted with two- and three-layer models

**Jodee Frias,*  Giles Blaney, Angelo Sassaroli, and Sergio Fantini**
Tufts University, Department of Biomedical Engineering, Medford, MA, USA

**Abstract**

**Significance:** Functional near-infrared spectroscopy (fNIRS) is impacted by signal contamination from superficial hemodynamics. It is important to develop methods that account for such contamination and provide accurate measurements of cerebral hemodynamics.

**Aim:** This work aims to investigate whether simulated data with two-layer or three-layer tissue models are able to reproduce *in vivo* data collected with dual-slope (DS) frequency-domain (FD) near-infrared spectroscopy (NIRS) on human subjects during brain activation.

**Approach:** We performed Monte Carlo simulations to generate DS FD-NIRS data from two- and three-layer media with a range of layer thicknesses and optical properties. We collected *in vivo* data with DS FD-NIRS (source-detector distances: 25, 37 mm; wavelengths: 690, 830 nm; modulation frequency: 140 MHz) over the occipital lobe of human subjects during visual stimulation. Simulated and *in vivo* data were analyzed with diffusion theory for a homogeneous medium and results were compared for each DS FD-NIRS data type.

**Results:** We found that the main qualitative features of *in vivo* data could be reproduced by simulated data from a three-layer medium, with a second layer (representing the cerebrospinal fluid in the subarachnoid space) that is less absorbing and less scattering than the other two layers, and with a top layer thickness that represents the combined scalp and skull thickness.

**Conclusions:** A three-layer model is a viable improvement over a homogeneous model to analyze DS FD-NIRS data (or any other fNIRS data) to generate more accurate measurements of cerebral hemodynamic changes without a need for large data sets for tomographic reconstructions.

*Corresponding Author, E-mail: Jodee.Frias@tufts.edu

**Statement of Discovery:** This study validates the use of a three-layer tissue model for analyzing functional dual-slope frequency-domain near-infrared spectroscopy (DS FD-NIRS) data, successfully reproducing the main qualitative features of *in vivo* functional data.

## 1 Introduction

Functional near-infrared spectroscopy (fNIRS) is a useful non-invasive tool for assessing brain activation through neurovascular coupling; however, the technique is impacted by a significant sensitivity to hemodynamic and oxygenation changes in superficial extracerebral tissue.[1] This prevailing problem complicates the method's ability to isolate the cerebral hemodynamic response to functional activation, as photons must travel through extracerebral layers (scalp, skull, cerebrospinal fluid (CSF), etc.) before reaching the cerebral tissue and then making it back to the tissue surface for detection.[1–4] This extracerebral contamination of NIRS signals is most prevalent in the simplest implementation of single-distance (one source and one detector) continuous-wave (CW) NIRS, which relies on measuring optical intensity changes and associated changes in the tissue absorption coefficient ($\Delta\mu_a$) and in the oxy- and deoxy-hemoglobin concentrations ($\Delta[HbO_2], \Delta[Hb]$).[5]

Tomographic approaches[6] and high-density diffuse optical tomography (HD-DOT)[7] are capable of spatially reconstructing the tissue hemodynamics within the probed tissue, thus separately assessing hemodynamics in superficial and deeper tissue.[8] This tomographic approach to fNIRS has demonstrated strong correlations with functional magnetic resonance imaging (fMRI), proving that non-invasive optical methods can achieve high-fidelity functional brain mapping when optical signals are properly analyzed to identify deeper cortical hemodynamics.[9] Despite the power of tomographic approaches, some of their features and requirements may limit their applicability and effectiveness in some scenarios. For example, the application of HD-DOT requires large numbers of sources and detectors that introduce physical and practical limitations, and uses computational methods that benefit from individual MRI images to generate subject-specific head models that may not always be available.[9] There is value in the development of simpler methods, from both instrumental and computational viewpoints, which can identify cerebral hemodynamics by suppressing or accounting for superficial hemodynamic contributions to the optical signals.

Many efforts have focused on discriminating extracerebral and cerebral contributions to optical signals without the use of a large number of sources and detectors or tomographic approaches. The most common approach uses multi-distance sets that include short source-detector distances ($\rho$) ($\rho < 10$ mm) to measure superficial tissue, and long source-detector distances ($\rho > 20$ mm) to measure a combination of superficial and deeper tissue. These measurements are then combined according to various data processing methods (least-squares fitting,[10] adaptive filtering,[11,12] independent component analysis (ICA)[13], etc.) to separate the hemodynamic contributions from superficial and deeper layers. While these methods have shown their effectiveness, they also have their own sets of drawbacks: least-squares regression and adaptive filtering both assume that the changes in superficial and deep layers are statistically independent, which may not always be true due to task-evoked systemic changes, and ICA can be difficult to streamline for large amounts of data.[12,13] Furthermore, the collection of data at short source-detector distances may be impacted by specular reflections at the tissue surface when the optical probe is not in good contact with the scalp.[14]

Time-resolved methods in the frequency-domain (FD) or time-domain (TD) have also been proposed to discriminate superficial and cerebral hemodynamics by leveraging measurements of the moments of the time-of-flight distribution in TD[15] or the phase of photon-density waves ($\phi$) in FD.[8] An additional advantage of time-resolved measurements is that they allow for the determination of absolute optical properties [absorption coefficient ($\mu_a$) and reduced scattering coefficient ($\mu_s'$)], and subsequently the absolute concentrations of oxy-hemoglobin ([HbO$_2$]) and deoxy-hemoglobin ([Hb]). This determination of absolute optical properties is important for a more quantitative characterization of the probed tissue.

A more recent development to achieve a preferential depth sensitivity is based on a special geometrical configuration of two light sources and two optical detectors, which was proposed in both the frequency domain[16] and time domain[17]. This configuration allows for the measurement of two slopes (or ratios) of optical data (namely, the linearized reflectance intensity ($\ln \rho^2 I$) and the phase ($\phi$) in the FD case) vs. $\rho$, which are then averaged to obtain a so-called dual-slope (DS), or dual-ratio, measurement. This geometrical configuration of sources and detectors employs two "long" distances for collecting data that are sensitive, to different extents, to both cerebral and extracerebral tissue. The absorption sensitivity of dual-slope measurements is preferentially deeper than that of individual single-distance measurements (especially for $\phi$ measurements in FD and for higher moments of the time-of-flight distribution in TD), suppressing contributions from superficial tissue.[18,19] A question, however, concerns the extent of residual extracerebral contributions to dual-slope measurements, and the impact of the layered tissue structure (scalp, skull, brain cortex, etc.) on the depth sensitivity of single-distance and dual-slope data. This question is investigated in this work.

While dual-slope data features a preferential deep sensitivity, translating measured changes of dual-slope data (say, intensity or phase in FD) into changes of tissue absorption or hemoglobin concentration requires a model to describe light propagation in tissue. A common option is diffusion theory for a semi-infinite homogenous medium, which provides a straightforward analytical method but raises questions on the reliability of its results given the oversimplification of the heterogenous anatomical structure of the human head. While the DS approach does suppress superficial confounds, some residual contributions from superficial layers cannot be completely avoided and they are expected to impact both the collected data and the accuracy of the homogeneous model used to obtain hemoglobin concentration dynamics. To robustly interpret the results obtained with DS FD-NIRS data analyzed with the homogenous diffusion model, it is important to understand how the underlying layered tissue influences the measured signals. This work investigates this point by modeling tissue as a two- or three-layered medium and also explores how these layered models are able to reproduce *in vivo* data measured with DS FD-NIRS on the occipital lobe of human subjects during visual stimulation.

Here, we report Monte Carlo simulations of light propagation in two- and three-layered scattering media to characterize how each DS FD-NIRS data type responds to simulated functional activation in the deepest layer, and to compare these simulation results to data collected *in vivo* during functional activation in human subjects with different extracerebral tissue thicknesses. The ultimate goal of this work is to identify a relatively simple tissue model (two- or three-layered medium) that can reproduce, at least qualitatively, the behavior of DS FD-NIRS data collected *in vivo* and that can serve as a basis for a more accurate determination of cerebral hemodynamics, improving upon the oversimplistic homogeneous tissue model. While a two- or three-layer model is still a simplification of the actual head anatomy, as long as it is able to reproduce the main features of *in vivo* data, it can still serve as a valuable tool for robust non-invasive measurements of cerebral hemodynamics without the need for large optical data sets or sophisticated personalized imaging for *a priori* information.

## 2 Methods

*2.1 Data collection*

*2.1.1. In-silico simulations of two- and three-layered media*

The first section of this work focuses on a comparison among the effectively homogeneous absorption changes obtained from each DS FD-NIRS data-type [namely, single-distance (SD) intensity ($I$) at $\rho$=25 mm (SD$I_{25\text{ mm}}$), SD$I$ at $\rho$=37 mm (SD$I_{37\text{ mm}}$), SD phase ($\phi$) at $\rho$=25 mm (SD$\phi_{25\text{ mm}}$), SD$\phi$ at $\rho$=37 mm (SD$\phi_{37\text{ mm}}$), dual-slope (DS) intensity ($I$) (DS$I$), and DS phase ($\phi$) (DS$\phi$)] in the case of absorption changes in the lowest layer of two- and three-layered media that mimic cerebral hemodynamics. These DS FD-NIRS data-types will be referred to as $M$ (for measured data) throughout this work, so that the effectively homogeneous absorption change obtained from each data-type is indicated by $\Delta\mu_{a,M}$. Due to the uncertainty surrounding *in vivo* absolute optical properties of the relevant tissue layers (scalp, skull, CSF, gray matter, white matter, etc.), we consider a wide range of biologically-relevant optical properties in accordance with *in vivo* and *ex vivo* measurements of such tissue layers, as shown in Figure 1.[3,4,7,20–23]

This work presents two different sets of simulations for both cases of two- and three-layered media. For two-layered media, the first set involves holding the $\mu_a$ value of each layer constant ($\mu_{a,1}$ for the first layer representing the scalp/skull and $\mu_{a,2}$ for the second layer representing the cerebral tissue), while the top-layer thickness ($L_1$) was varied and $\mu'_{s,1}/\mu'_{s,2}$ was varied by fixing $\mu'_{s,2}$ and varying $\mu'_{s,1}$ ($\mu'_{s,1}$ for the first layer representing the scalp/skull and $\mu'_{s,2}$ for the second layer representing the cerebral tissue). The second set involves holding the $\mu'_s$ value of each layer constant, while the top-layer thickness ($L_1$) was varied and $\mu_{a,1}/\mu_{a,2}$ was varied by fixing $\mu_{a,2}$ and varying $\mu_{a,1}$. This allowed for the investigation of the influence of the values of $\mu_a$ and $\mu'_s$ of each layer, the relative absorption and scattering properties of the two layers ($\mu_{a,1}/\mu_{a,2}$ and $\mu'_{s,1}/\mu'_{s,2}$), and $L_1$ on the recovered $\Delta\mu_{a,M}$ associated with absorption changes in the bottom layer ($\Delta\mu_{a,2}$). For three-layered media, the same procedure was followed, except an additional layer was added, creating a structure where the first layer represents scalp/skull, the second layer represents the CSF-filled subarachnoid space, and the third layer represents the cerebral tissue. The optical properties of the second layer representing the CSF were held constant and assigned in accordance with explored values in literature (see also Section 4.2) [3,4,22,23], to investigate the influence of the presence of a low-absorption/low-scattering layer representing the CSF on the recovered $\Delta\mu_{a,M}$ associated with absorption changes in the bottom layer ($\Delta\mu_{a,3}$). The optical properties of the first and third layer were then varied in the same manner as those of the first and second layer in the two-layer medium, and $L_1$ was incremented stepwise to simulate the varying scalp/skull thicknesses of adult human heads. Relevant optical properties and layer thicknesses for all simulation sets (two- and three-layer) are reported in Figure 1.[3,4,7,20–23]

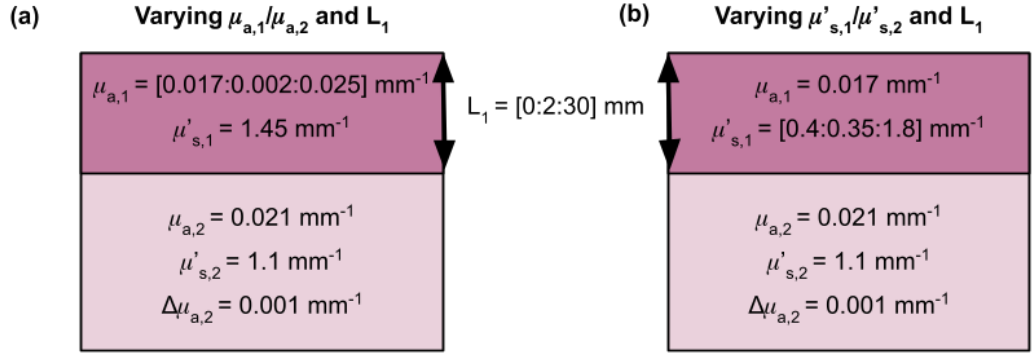

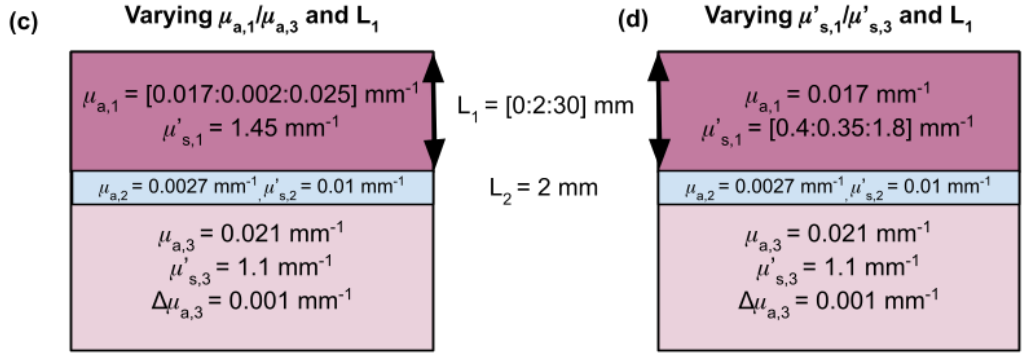

Figure 1. Layered tissue models used for Monte Carlo simulations in Monte Carlo eXtreme (MCX). Two-layer models (top) and three-layer models (bottom) are shown, illustrating variations in the optical properties where $\mu'_s$ is varied in the left column, and $\mu_a$ is varied in the right column. Notation of [X:Y:Z] means that X is the minimum value, Y is the step size of the iteration of said parameter, and Z is the maximum value.

All simulations were conducted using Monte Carlo eXtreme (MCX)[24] with an overall rectangular prism geometry of [200×100×100] mm³. An optode arrangement allowing for the utilization of the dual-slope configuration ($\rho$=[25, 37] mm) was applied to the top surface of the medium, with an optical detector radius of 1 mm and a pencil beam source. The anisotropy factor ($g$-value) was set to 0, and the index of refraction was set to 1.4 for each tissue layer and 1.0 for the outside medium. $9 \times 10^9$ photons were launched for each simulation. These MCX simulations were run for zero absorption to find the value of $\ell_{j,k}$, the path length of the $k$th photon in layer $j$, and the absorption of each layer is introduced to find the complex reflectance ($\tilde{R}$), as follows:

$$\tilde{R} = \frac{1}{N_{\text{detp}} A_{\text{det}}} \sum_{k=1}^{N_{\text{detp}}} e^{-\sum_{j=1}^{N_l} \mu_{a,j} \ell_{j,k}} * e^{i \sum_{j=1}^{N_l} \omega \frac{\ell_{j,k}}{c/n_j}} \qquad (1)$$

where $N_{\text{detp}}$ is the number of detected photons, $A_{\text{det}}$ is the area of the detector, $N_l$ is the number of layers in the medium, $\mu_{a,j}$ is the absorption coefficient of layer $j$, $c$ is the speed of light in a vacuum, $n_j$ is the refractive index of layer $j$, and $\omega$ is the angular modulation frequency. To simulate functional brain activation, all simulations included an absorption change in the deepest layer ($\Delta\mu_{a,2}$ for two-layer simulations, $\Delta\mu_{a,3}$ for three-layer simulations) of 0.001 mm$^{-1}$,[25–27] and the calculation of the associated change in the complex reflectance ($\Delta\tilde{R}$). These $\Delta\tilde{R}$ values were then analyzed using the same approach as the experimental data, as described in Section 2.2, to retrieve values of $\Delta\mu_{a,M}$ that can be compared with those obtained from the *in vivo* measurements.

### 2.1.2. In vivo brain measurements

*Equipment and human subjects*

*In-vivo* brain measurements were obtained with the ISS Imagent V2 (ISS, Champaign, IL, USA) FD-NIRS instrument. This instrument utilizes laser diodes that emit light at two wavelengths ($\lambda$s) of 690 and 830 nm and that are intensity-modulated at a frequency of 140.625 MHz. Optical fibers connected the ISS Imagent V2 to two modular hexagonal dual-slope arrays[28–30], allowing for bilateral measurements on the human occipital lobe. Each module included two optical detectors and four dual-wavelength optical sources, forming four DS sets and eight SD sets in each module, with one DS set spanning the two modules. The setup and modules are shown in Fig. 2 and more details can be found in Ref. [28]. The probes were placed bilaterally over the occipital region above the primary visual cortex. The size of each module is approximately 40 mm × 40 mm, allowing for tessellation of modules over the primary visual cortex.[28] In this work, we recruited three healthy adult subjects (ages 25-27, 2 female) for our Tufts University Institutional Review Board (IRB) approved functional brain activation protocol. Following data acquisition, each subject had their combined scalp and skull thickness (represented by $L_1$ in the simulations) measured with ultrasound imaging (SonoSite S-Nerve FUJIFILM, SonoSite, Inc., Bothell, WA, USA). To determine the total extracerebral thickness, specific regions of interest (ROI) were identified within the ultrasound images. The scalp thickness was measured as the distance between the skin surface and the upper reflection of the skull, and the skull thickness was measured as the distance between the upper and lower reflections from the skull after taking into account the speed of ultrasound in scalp or skull tissue.[31]

*Visual stimulation*

The *in vivo* visual stimulation protocol consisted of an initial baseline at rest (1 min), 7 repetitions of stimuli (15 s) and rest (30 s) periods, and a final baseline at rest (1 min).[28–30] The initial baseline was used for absolute optical property retrieval, further explained in Section 2.2.1. The stimulation consisted of a 62-cm diameter, 8-Hz contrast reversing checkerboard, displayed on an eye-level screen 1.8 m away from the seated subject.[32] This stimulation is designed to elicit activation across the primary visual cortex in a block-protocol design, to then allow for a folding average to be taken across stimulus-rest periods to increase signal-to-noise ratio (SNR), as elaborated on in Section 2.2.3.

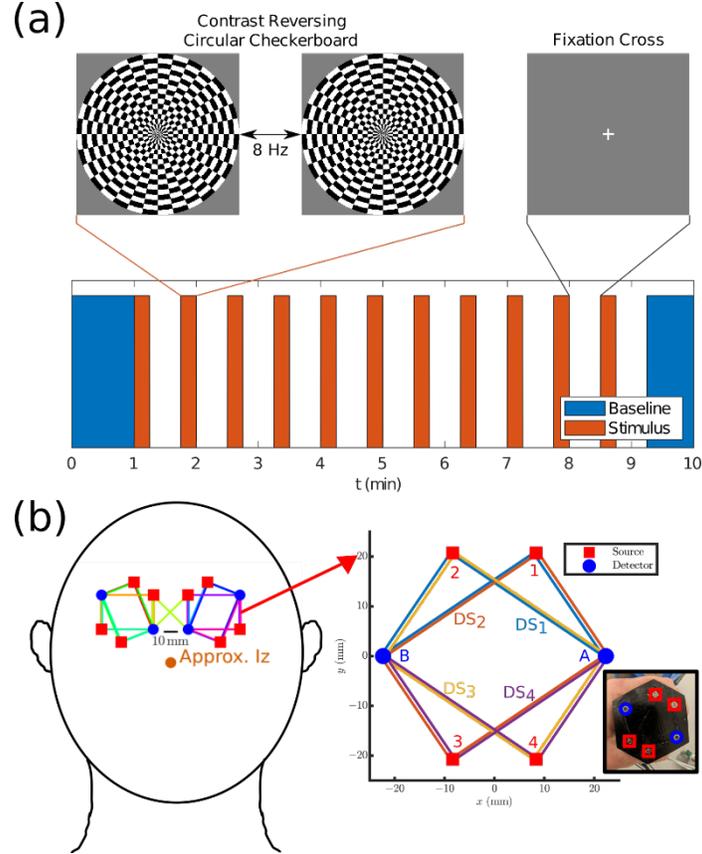

Figure 2. (a) Visual stimulation paradigm initiated over time of measurement. Orange periods indicate stimulus, where the contrast reversing checkerboard is enacted. Blue periods indicate baseline periods used to retrieve absolute optical properties. White regions indicate rest, where a fixation cross is shown. (b) Placement of hexagonal probes on the head of the subject, where each hexagonal probe contains four dual-slope sets and eight single-distance sets. The probes are placed approximately 10 mm apart above the approximate location of the inion, over the occipital lobe (primary visual cortex) of the subject.

*2.2 Data analysis*

*2.2.1. Absolute optical properties ($\mu_a, \mu_s'$)*

In order to conduct measurements of the effective absorption changes (i.e. the absorption changes resulting from an assumption that the tissue is homogeneous) associated with data-type $M$ ($\Delta\mu_{a,M}$), the effectively homogenous optical properties ($\mu_a, \mu_s'$) were obtained at the initial baseline using the self-calibrated DS FD-NIRS data analyzed with diffusion theory for a semi-infinite homogenous medium with extrapolated boundary conditions.[16,33] Briefly, this method uses the $\tilde{R}$ vs. $\rho$ dependence obtained from self-calibrated data for each DS set to determine an initial guess of the $\mu_a$ and $\mu_s'$ values that are then iteratively updated to find the optical properties that are consistent with the baseline data.[33] For *in silico* simulations, baseline data are obtained from the complex reflectance ($\tilde{R}$) prior to the incorporation of a $\Delta\mu_a$ perturbation in the deepest layer. For *in vivo* experiments, the data collected during the 60 s baseline period prior to functional simulation are used to recover a $\mu_a$ and $\mu_s'$ for each DS set, which were then averaged across the spatial region of the analyzed hexagonal module to obtain one set of optical properties for each subject.

*2.2.2. Changes in absorption coefficient ($\Delta\mu_{a,M}$)*

Recovered effective $\mu_a$ and $\mu_s'$ (as described in Section 2.2.1) were then used to obtain the partial derivative of a given $M$ versus $\mu_a$ ($J_{M,\mu_a}$) for a homogeneous medium that has the given absolute $\mu_a$ and $\mu_s'$. The recovered $\Delta\mu_{a,M}$ is then given by:

$$\Delta\mu_{a,M} = \frac{\Delta M}{J_{M,\mu_a}}, \qquad (2)$$

where $M$ can be $\ln(\rho^2 I)$, $\Delta\ln(\rho^2 I)/\Delta\rho$, $\phi$, or $\Delta\phi/\Delta\rho$ collected in SD or DS configuration (i.e., SD$I$, SD$\phi$, DS$I$, DS$\phi$). These $\Delta\mu_{a,M}$ values at two wavelengths can then be converted into effectively homogeneous changes in oxy-hemoglobin concentration ($\Delta[HbO_2]$) and deoxy-hemoglobin concentration ($\Delta[Hb]$) using Beer's law.[34]

*2.2.3. Measure of functional brain activation with data-type M (maximum $\Delta\mu_{a,M}$ at 830 nm)*

For the *in vivo* functional data, the $\Delta\mu_{a,M}$ values collected over the entire protocol duration were subjected to a folding-average over the 45 s stimulus-rest cycles. This yielded one representative functional activation trace for each data-type and measurement set (e.g., 1A for SD data obtained with source 1 and detector A, 1AB2 for DS data obtained with the dual slope set that comprises sources 1 and 2, and detectors A and B). This step enhances the SNR while minimizing systemic contributions and motion artifacts that are not synchronous with the cyclic visual stimulation.

Data analysis focused on one module, specifically the one positioned over the left primary occipital cortex due to a better optical coupling and SNR when compared to the second module. This analyzed module contained four-dual slope sets and their eight constituent single-distance sets (four $\rho$=25 mm sets, four $\rho$=37 mm sets). A mapping algorithm was utilized to identify and group the specific SD source-detector pairs that geometrically correspond to each DS set. Prior to peak detection, the time-courses of these paired SD channels (such as the two $\rho$=25 mm sets compromising a specific DS configuration) were averaged. The primary parameter utilized for subsequent analysis was the maximum $\Delta\mu_{a,M}$ at 830 nm, as a positive $\Delta\mu_{a,M}$ at 830 nm could be associated with an increase in blood flow (positive $\Delta[HbO_2]$, negative $\Delta[Hb]$) in response to the increased metabolic demand elicited by functional activation.[31,33] This allows us to simplify our analysis and represent the hemodynamic response to brain activation measured with each data-type *M* with a single parameter. For *in vivo* data, a custom MATLAB function was developed for peak $\Delta\mu_{a,M}$ at 830 nm determination. The folding-averaged signal was first smoothed using a moving average window of 3 s. The standard deviation ($\sigma$) of the signal over an intermediate 2 s segment was utilized for noise estimation. A detection threshold of $3\sigma$ was applied to ensure all detected maximum $\Delta\mu_{a,M}$ values represented a functional absorption increase above the noise level. The time corresponding to the maximum value was identified, and the final reported maximum $\Delta\mu_{a,M}$ was calculated as the mean value within a $\pm 1$ second window centered on the detected peak time. The resulting maximum $\Delta\mu_{a,M}$ values were averaged across all $M$ data collected with the entire left-side module. The error is represented by the standard error of the mean (SEM) for each $M$. For *in silico* simulation data, maximum $\Delta\mu_{a,M}$ values were defined consistently with the *in vivo* definitions for each $M$.

## 3. Results

### 3.1. Monte Carlo simulations of DS FD-NIRS data-types and associated $\Delta\mu_{a,M}$

*3.1.1 Two-layer Monte Carlo simulations: varying $\mu_a$*

Figure 3(a) shows the $\Delta\mu_{a,M}$ values associated with DS FD-NIRS data-types that were generated with Monte Carlo simulations of two-layered media with a functional absorption change in the bottom (second) layer, as a function of superficial layer thickness ($L_1$) and for different absorption coefficients of the top layer ($\mu_{a,1}$) (detailed in Figure 1(a)). As expected, increasing $L_1$ results in a decrease of all $\Delta\mu_{a,M}$ reflecting the reduced sensitivity to the bottom layer as it gets deeper, with the exception of $M = \text{DS}\phi$ which undergoes a minor increase prior to its decrease starting at $L_1=10$ mm. Notably, while varying $\mu_{a,1}$ scaled the absolute value of the recovered $\Delta\mu_{a,M}$ values, it did not alter the relative sensitivity hierarchy of the data-types, starting from $L_1=10$ mm. Across these $L_1$ values, the relative hierarchy of the data-types follows this order: $\Delta\mu_{a,\text{DS}\phi} > \Delta\mu_{a,\text{SD}\phi_{37\text{ mm}}} > \Delta\mu_{a,\text{DS}I} > \Delta\mu_{a,\text{SD}\phi_{25\text{ mm}}} > \Delta\mu_{a,\text{SD}I_{37\text{ mm}}} > \Delta\mu_{a,\text{SD}I_{25\text{ mm}}}$, in the case of these specific scattering properties. Across all $L_1$, $\Delta\mu_{a,\text{DS}\phi}$ is the measured absorption change that is the closest to the perturbation induced in the deepest layer (simulating functional activation). Figure 3(b)-(d) show the cases of three specific values of $L_1$ (10, 12, and 14 mm), which are chosen as being representative of the $L_1$ values measured *in vivo* across three subjects using ultrasound imaging (see Table 1). Figures 3(b)-(d) illustrate the impact of superficial tissue absorption at these fixed $L_1$s, with the x-axis reporting each $M$. It is observed that across these three fixed $L_1$s, a larger $\mu_{a,1}$ leads to a larger retrieved $\Delta\mu_{a,M}$ for all $M$s. Additionally, it is observed that across both $I$ and $\phi$ data-types, $\Delta\mu_{a,M}$ increases from $SD_{25\text{ mm}}$ to $SD_{37\text{ mm}}$ to $DS$.

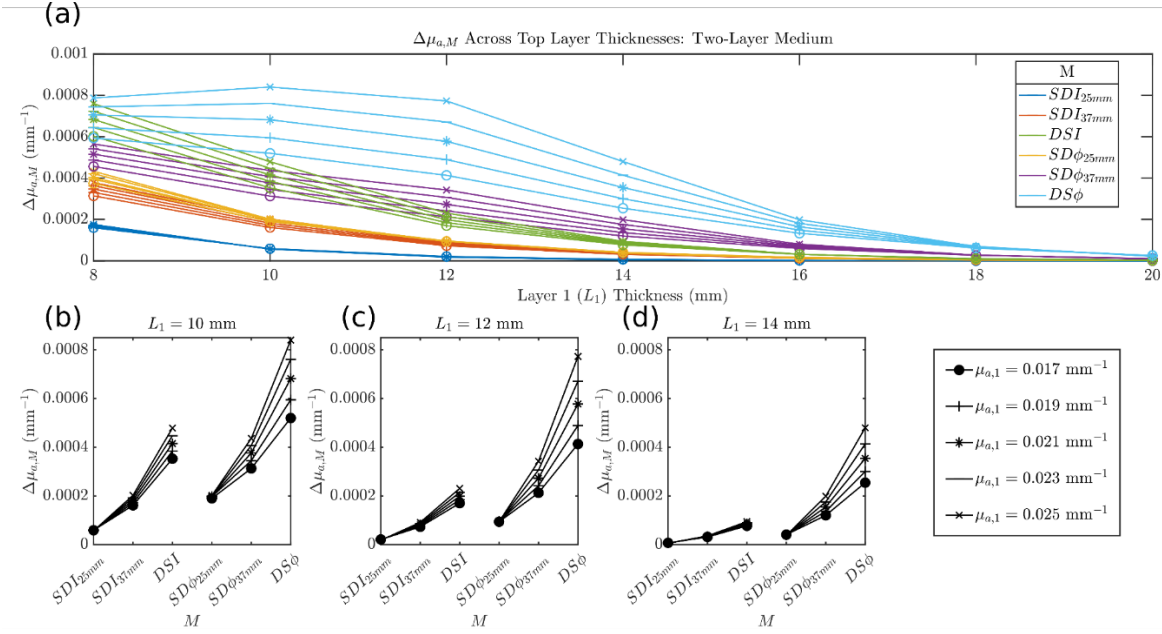

Figure 3. Evaluation of $\Delta\mu_{a,M}$ where $M$ can be $I$ or $\phi$ collected in SD or DS configuration (i.e., SD$I$, SD$\phi$, DS$I$, DS$\phi$) of a two-layer medium where $\mu_{a,1}$ is varied and all other optical properties are held constant. The top panel shows $\Delta\mu_{a,M}$ as a function of top layer thickness ($L_1$), while the bottom panels detail this relationship for specific $L_1$ values (10 mm, 12 mm, 14 mm). Properties held constant: $\Delta\mu_{a,2}$=0.001 mm$^{-1}$, $\mu'_{s,1}$ = 1.45 mm$^{-1}$, $\mu'_{s,2}$ = 1.1 mm$^{-1}$, $\mu_{a,2}$ = 0.021 mm$^{-1}$.

*3.1.2. Two-layer Monte Carlo simulations: varying $\mu'_s$*

In contrast to the effect of varying absorption, changing the superficial scattering coefficient ($\mu'_s$) significantly altered the relative hierarchy of the data-types. Figure 4(a) depicts the influence of superficial tissue scattering ($\mu'_{s,1}$) and $L_1$ on $\Delta\mu_{a,M}$ for a simulated functional absorption change in the bottom layer of a two-layer tissue model (detailed in Figure 1(b)). Similarly to Figure 3(a), we observe a decrease of all $\Delta\mu_{a,M}$ as $L_1$ increases. This decrease is less monotonic and uniform than that observed in Figure 3, which becomes more apparent when analyzing Figure 4(b)-(d). The first observation of note in Figure 4(b)-(d) is that in contrast to Figure 3 where $\mu_{a,1}$ was varied and the order of retrieved $\Delta\mu_{a,M}$ values for $L_1 \geq 10$ mm ($\Delta\mu_{a,DS\phi} > \Delta\mu_{a,SD\phi_{37\,mm}} > \Delta\mu_{a,DSI} > \Delta\mu_{a,SD\phi_{25\,mm}} > \Delta\mu_{a,SDI_{37\,mm}} > \Delta\mu_{a,SDI_{25\,mm}}$) stayed the same despite the variation of $\mu_{a,1}$, this is not the case for varying $\mu'_{s,1}$. Looking at Figure 4(b) where $L_1 = 10$ mm, when $\mu'_{s,1}/\mu'_{s,2}<1$, the order of the retrieved $\Delta\mu_{a,M}$ values is as follows: ($\Delta\mu_{a,DS\phi} > \Delta\mu_{a,SD\phi_{37\,mm}} > \Delta\mu_{a,SD\phi_{25\,mm}} > \Delta\mu_{a,DSI} > \Delta\mu_{a,SDI_{37\,mm}} > \Delta\mu_{a,SDI_{25\,mm}}$). When $\mu'_{s,1}/\mu'_{s,2}>1$, the order changes to: ($\Delta\mu_{a,DS\phi} > \Delta\mu_{a,SD\phi_{37\,mm}} > \Delta\mu_{a,DSI} > \Delta\mu_{a,SD\phi_{25\,mm}} > \Delta\mu_{a,SDI_{37\,mm}} > \Delta\mu_{a,SDI_{25\,mm}}$). This order switch when $\mu'_{s,1}/\mu'_{s,2}>1$ (namely $\Delta\mu_{a,DSI} > \Delta\mu_{a,SD\phi_{25\,mm}}$) is also observed in Figure 4(c) where $L_1 = 12$ mm and Figure 4(d) where $L_1 = 14$ mm, although the values of $\Delta\mu_{a,DSI}$ and $\Delta\mu_{a,SD\phi_{25\,mm}}$ tend to get closer to each other. This "crossover" effect between $\Delta\mu_{a,DSI}$ and $\Delta\mu_{a,SD\phi_{25\,mm}}$ illustrates that the relative depth sensitivity of data-types is dependent upon the scattering contrast between tissue layers.

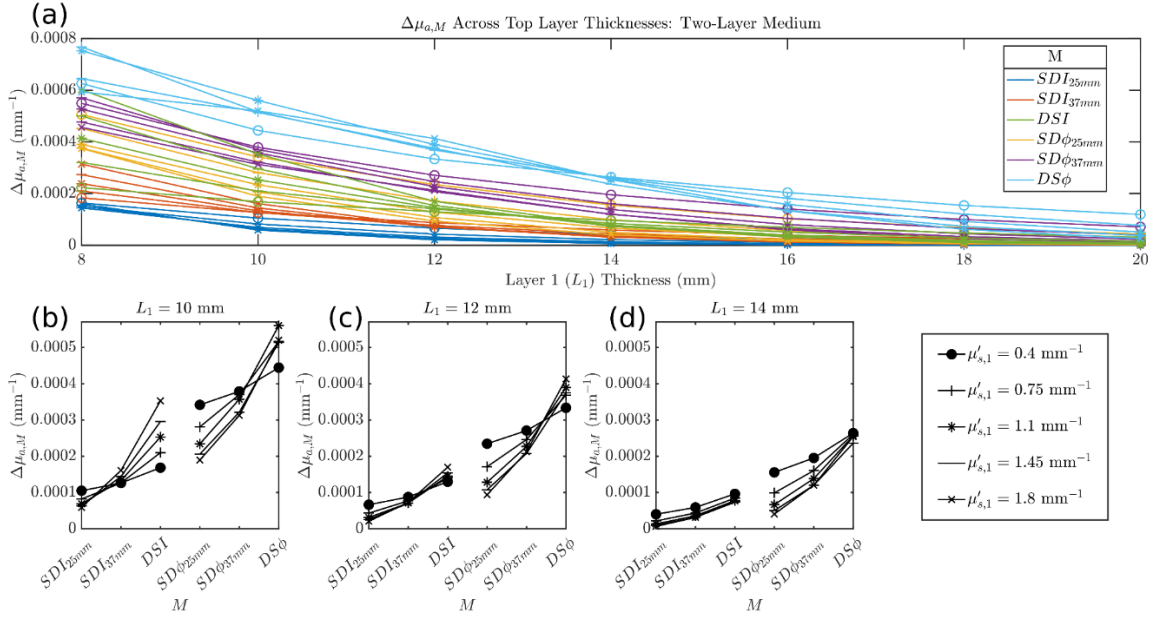

Figure 4. Evaluation of $\Delta\mu_{a,M}$ where $M$ can be $I$ or $\phi$ collected in SD or DS configuration (i.e., SD$I$, SD$\phi$, DS$I$, DS$\phi$) of a two-layer medium where $\mu'_{s,1}$ is varied and all other optical properties are held constant. The top panel shows $\Delta\mu_{a,M}$ as a function of top layer thickness ($L_1$), while the bottom panels detail this relationship for specific $L_1$ values (10 mm, 12 mm, 14 mm). Properties held constant: $\Delta\mu_{a,2}$=0.001 mm$^{-1}$, $\mu_{a,1}$ = 0.017 mm$^{-1}$, $\mu_{a,2}$ = 0.021 mm$^{-1}$, $\mu'_{s,2}$ = 1.1 mm$^{-1}$.

### 3.1.3 Three-layer Monte Carlo simulations: varying $\mu_a$

Figure 5 investigates the retrieved $\Delta\mu_{a,M}$ from a three-layer medium with an absorption change in the deepest (third) layer, focusing on how the measurement is affected by changes in $L_1$ and $\mu_{a,1}$ (detailed in Figure 1(c)). This medium includes a CSF-mimicking tissue layer, whose thickness is held constant at $L_2$=2 mm. As in Figures 3 and 4, we see a decrease in $\Delta\mu_{a,M}$ as $L_1$ increases for all $M$, but there is the notable exception of $\Delta\mu_{a,DSI}$, which increases between $L_1$=8 and 10 mm before decreasing starting at $L_1$=10 mm. These observed decreases in $\Delta\mu_{a,M}$ are less pronounced than those observed in the two-layer medium. Additionally, it can be seen that while with the two-layer medium we observed a consistent order of $\Delta\mu_{a,M}$ values from greatest to smallest for each $L_1$, this consistency is not maintained for the three-layer medium. This is shown at $L_1$=18 mm for example, where $\Delta\mu_{a,\text{DS}I} < \Delta\mu_{a,\text{SD}\phi 37\text{ mm}}$ while the opposite is true at $L_1$=12 mm. This fact is maintained, however, regardless of the value of $\mu_{a,1}$ and seems to be solely dependent on $L_1$. Figure 5(b)-(d) show three specific values of $L_1$ (10, 12, and 14 mm), where one can appreciate the impact of superficial baseline absorption at these fixed $L_1$s, with the x-axis reporting each $M$. It is observed that across all of these fixed $L_1$s, a larger $\mu_{a,1}$ leads to a larger retrieved $\Delta\mu_{a,M}$ for all $M$s. This being said, the $\mu_{a,1}$ values considered here do not seem to have an effect on the order of the retrieved $\Delta\mu_{a,M}$.

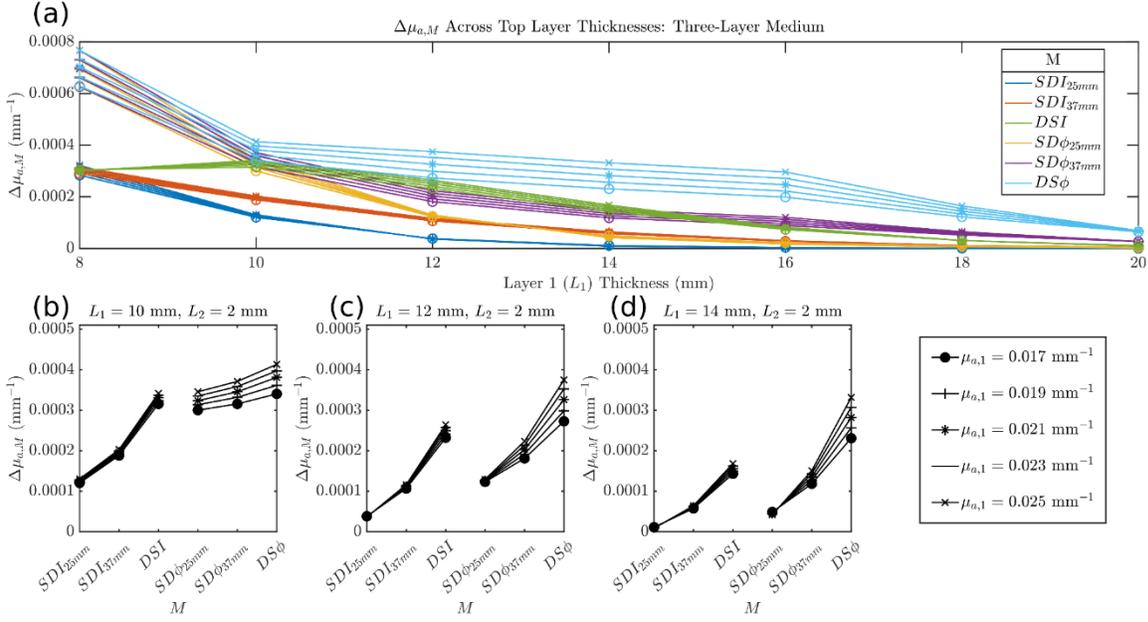

Figure 5. Evaluation of $\Delta\mu_{a,M}$ where $M$ can be $I$ or $\phi$ collected in SD or DS configuration (i.e., SD$I$, SD$\phi$, DS$I$, DS$\phi$) of a three-layer medium where $\mu_{a,1}$ is varied and all other optical properties are held constant. The top panel shows $\Delta\mu_{a,M}$ as a function of top layer thickness ($L_1$), while the bottom panels detail this relationship for specific $L_1$ values (10 mm, 12 mm, 14 mm). Properties held constant: $\Delta\mu_{a,3}$=0.001 mm$^{-1}$, $\mu'_{s,1}$ = 1.45 mm$^{-1}$, $\mu'_{s,2}$ = 0.01 mm$^{-1}$, $\mu'_{s,3}$ = 1.1 mm$^{-1}$, $\mu_{a,2}$ = 0.0027 mm$^{-1}$, $\mu_{a,3}$ = 0.021 mm$^{-1}$.

## 3.1.4. Three-layer Monte Carlo simulations: varying $\mu'_s$

Figure 6 investigates the retrieved $\Delta\mu_{a,M}$ in a three-layer medium with an absorption perturbation in the deepest (third) layer, focusing on changes in $L_1$ and $\mu'_{s,1}$ in the presence of a second CSF-mimicking tissue layer (shown in Figure 1(d)). In agreement with Figure 5, we observe a decrease in $\Delta\mu_{a,M}$ as $L_1$ increases for all $M$ except for DS$I$, where $\Delta\mu_{a,DSI}$ increases below $L_1=10$ mm and decreases starting at $L_1=10$ mm. Additionally, there is the same $L_1$-dependent trend of the order of $\Delta\mu_{a,M}$ values as in Figure 5. There is also a similar trend as a function of $\mu'_{s,1}$ that was observed in Figure 4. Here, when looking at Figure 6(b) ($L_1 = 10$ mm), when $\mu'_{s,1}/\mu'_{s,2}<1$, the order of the $\Delta\mu_{a,M}$ values is as follows: ($\Delta\mu_{a,DS\phi} > \Delta\mu_{a,SD\phi 37\text{ mm}} > \Delta\mu_{a,SD\phi 25\text{ mm}} > \Delta\mu_{a,DSI} > \Delta\mu_{a,SDI 37\text{ mm}} > \Delta\mu_{a,SDI 25\text{ mm}}$), whereas when $\mu'_{s,1}/\mu'_{s,2}>1$, it becomes: ($\Delta\mu_{a,DS\phi} > \Delta\mu_{a,SD\phi 37\text{ mm}} > \Delta\mu_{a,DSI} > \Delta\mu_{a,SD\phi 25\text{ mm}} > \Delta\mu_{a,SDI 37\text{ mm}} > \Delta\mu_{a,SDI 25\text{ mm}}$). We see that this trend becomes less consistent as $L_1$ is increased, most easily seen in Figure 6(d), where at $\mu'_{s,1} = 0.4$ mm$^{-1}$, $\Delta\mu_{a,DSI}$ is approximately equal to that $\Delta\mu_{a,SD\phi 25\text{ mm}}$. This illustrates an impact not only from $\mu'_{s,1}/\mu'_{s,2}$, but also from $L_1$.

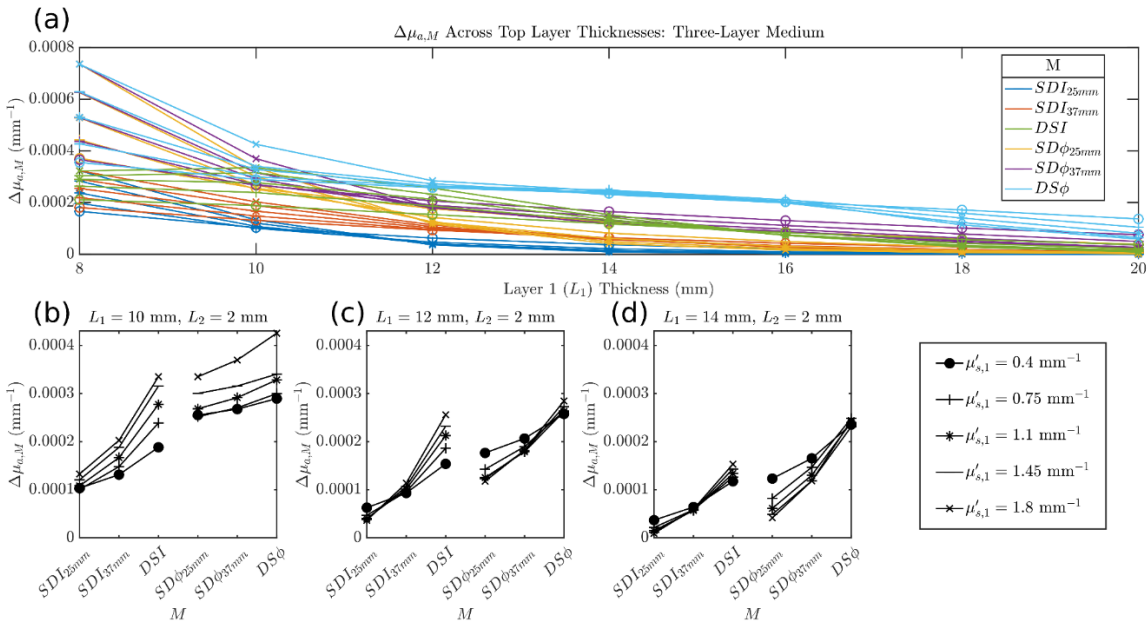
1919

Figure 6. Evaluation of $\Delta\mu_{a,M}$ where $M$ can be $I$ or $\phi$ collected in SD or DS configuration (i.e., SD$I$, SD$\phi$, DS$I$, DS$\phi$) of a three-layer medium where $\mu'_{s,1}$ is varied and all other optical properties are held constant. The top panel shows $\Delta\mu_{a,M}$ as a function of top layer thickness ($L_1$), while the bottom panels detail this relationship for specific $L_1$ values (10 mm, 12 mm, 14 mm). Properties held constant: $\Delta\mu_{a,3}$=0.001 mm$^{-1}$, $\mu_{a,1}$ = 0.017 mm$^{-1}$, $\mu_{a,2}$ = 0.0027 mm$^{-1}$, $\mu_{a,3}$ = 0.021 mm$^{-1}$, $\mu'_{s,2}$ = 0.01 mm$^{-1}$, $\mu'_{s,3}$ = 1.1 mm$^{-1}$.

## 3.2. Experimental results in-vivo

### 3.2.1. Representative time traces

Figure 7(a) reports the folding-average time traces of $\Delta[HbO_2]$ and $\Delta[Hb]$ during the 15 s stimulation and 30 s rest period obtained from Subject 1 for one DS set (DS Set 4 in Fig. 2(b)) and the SD sets that comprise it. Figure 7(b) reports the $\Delta\mu_{a,M}$ values at 830 nm for the same data sets of Figure 7(a). In the time traces of Fig. 7(a), there is a clear difference between the hemodynamic changes retrieved by each $M$, a trend which is preserved when observing the amplitude of $\Delta\mu_{a,M}$ at 830 nm in Fig. 7(b). From these traces, one can see that hemodynamic and absorption changes obtained with DS$\phi$ have the largest amplitude with respect to those obtained with all other data-types in this subject, indicating that DS$\phi$ is more preferentially sensitive to deeper (cerebral) tissue for this subject. Hemodynamic and absorption changes obtained with SD$\phi$ (both $\rho$=25 and 37 mm) then show the next highest amplitude, followed by DS$I$ and then the more conventional SD$I$ data-type (both $\rho$=25 and 37 mm). This finding elucidates the different depth sensitivities provided by each $M$, as more preferential sensitivity to deeper (cerebral) tissue results in a greater $\Delta\mu_{a,M}$, since for this functional task the hemodynamic response elicited in the deeper, cerebral tissue is expected to be significantly greater than that in the superficial, extracerebral tissue.[25] The increase in blood flow in response to functional activation results in an increase in [HbO$_2$] (positive

Δ[HbO$_2$]), a decrease in [Hb] (negative Δ[Hb]), and an increase in absorption at 830 nm (positive Δ$\mu_{a,M}$).[31,33]

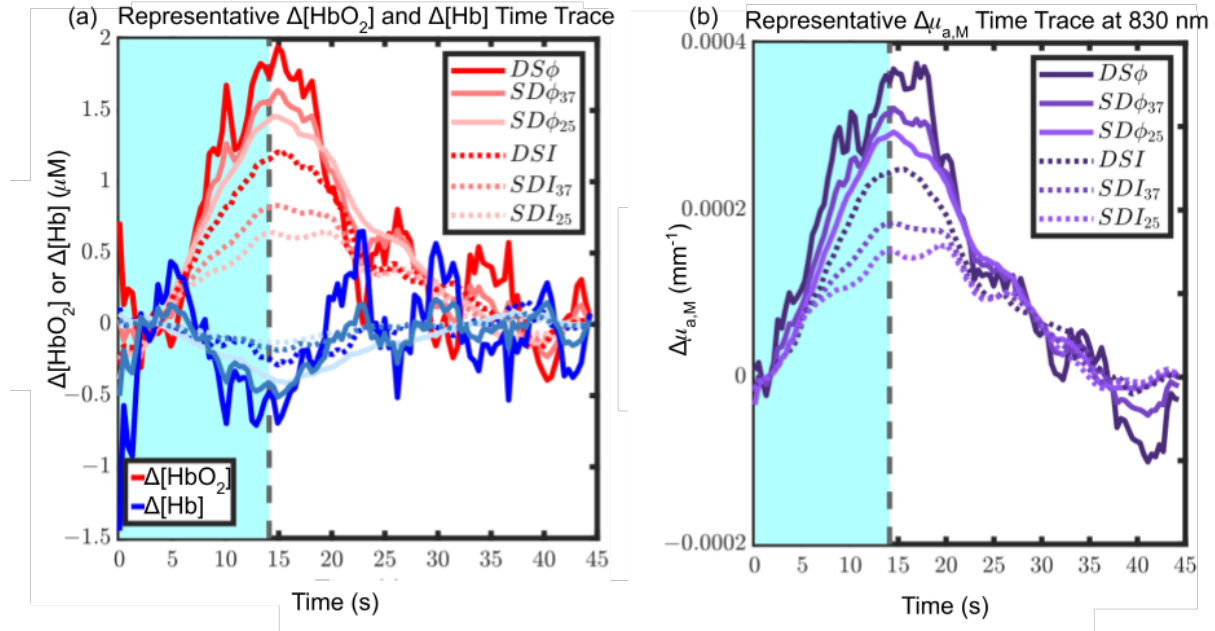

Figure 7. (a) Representative Δ[HbO$_2$]($t$) and Δ[Hb]($t$) folding-averaged time trace from Subject 1, where the visual contrast reversing checkerboard stimuli is enacted during the cyan region, and the region following the dotted line is the rest. (b) Representative Δ$\mu_{a,M}$($t$) folding-averaged time trace at 830 nm from Subject 1, corresponding to the trace in (a). It is visually evident that the hemodynamic trends observed in (a) are preserved when only considering Δ$\mu_{a,M}$ at 830 nm in (b). A moving-average filter with a window of 3 s has been applied to the time traces following the folding-average.

*3.2.3. Comparison between in-vivo and theoretical results*

Figure 8(a) reports the Δ$\mu_{a,M}$ at 830 nm obtained by data-type $M$ for all three subjects of increasing extracerebral (scalp and skull) tissue thickness (smallest for Subject 1 and greatest for Subject 3, as measured by ultrasound imaging described in Section 2.1.2., reported in Table 1). Subject 1 consistently exhibits the highest amplitude of $\Delta\mu_{a,M}$ for all $M$. As this subject has the thinnest superficial extracerebral tissue, this is likely due to a greater sensitivity to the shallower cortical tissue. Conversely, Subject 3 consistently exhibits the lowest $\Delta\mu_{a,M}$ magnitude as a result of the deepest cortical tissue among the three subjects.

One can also observe a consistent trend across all three subjects: for SD measurements (for both $I$ and $\phi$), $\Delta\mu_{a,M}$ increases when increasing $\rho$ from 25 to 37 mm. This is expected, as longer source-detector separations generally probe deeper tissue, where the functional absorption change is occurring. For all subjects, the DS configuration of both $I$ and $\phi$ yields a greater $\Delta\mu_{a,M}$ than the corresponding SD data at 37 mm (i.e. $\Delta\mu_{a,\mathrm{DS}I} > \Delta\mu_{a,\mathrm{SD}I_{37\,mm}}$ and $\Delta\mu_{a,\mathrm{DS}\phi} > \Delta\mu_{a,\mathrm{SD}\phi_{37\,mm}}$). This further confirms on human subjects *in vivo* the previously reported advantages of the DS data in terms of a preferential depth sensitivity compared to SD data.[16,33]

In Figure 8(a), one can notice a difference in the relative values of $\Delta\mu_{a,M}$ obtained with intensity and phase data types in the three subjects. Specifically, in Subject 1 (thinnest extracerebral tissue) $\Delta\mu_{a,\mathrm{SD}\phi} > \Delta\mu_{a,\mathrm{DS}I}$, in Subject 2 (intermediate thickness of extracerebral tissue) $\Delta\mu_{a,\mathrm{SD}\phi} \approx \Delta\mu_{a,\mathrm{DS}I}$, and in Subject 3 (thickest extracerebral tissue) $\Delta\mu_{a,\mathrm{SD}\phi} < \Delta\mu_{a,\mathrm{DS}I}$. This indicates a dependence of the relative brain sensitivity of the different DS FD-NIRS data-types on the thickness of the extracerebral tissue.

To investigate and better understand the origin of the *in vivo* results reported in Figure 8(a) for $\Delta\mu_{a,M}$, we show in Figure 8 (b), (c), and (d) the results of two- and three-layered medium simulations for a range of geometrical ($L_1$ values) and/or scattering conditions. Figure 8(b) reports $\Delta\mu_{a,M}$ for a two-layered medium with an absorption perturbation in the second layer, where $L_1$ and $\mu'_{s,2}$ are kept constant at 10 mm and 1.1 mm$^{-1}$, respectively, and $\mu'_{s,1}/\mu'_{s,2}$ takes values of 0.36 or 1.63 corresponding to $\mu'_{s,1} = 0.4$ mm$^{-1}$ (magenta) or $\mu'_{s,1} = 1.8$ mm$^{-1}$ (green). This simulation was conducted for other values of $L_1$ that drew the same conclusions. Varying $\mu'_{s,1}/\mu'_{s,2}$ in a two-layered medium recreates the opposing relationship between $\Delta\mu_{a,SD\phi}$ and $\Delta\mu_{a,DSI}$ observed between Subjects 1 and 3 *in vivo*, but fails to recreate the different $\Delta\mu_{a,M}$ magnitudes observed in Subjects 1 and 3, which is likely linked to different extracerebral tissue thicknesses in the two subjects. Figure 8(c) considers the case of different top layer thicknesses by reporting results on a two-layered medium where $L_1 = 10$ mm and $\mu'_{s,1} = 0.4$ mm$^{-1}$ (magenta) or $L_1 = 14$ mm and $\mu'_{s,1} = 1.8$ mm$^{-1}$ (green) (with $\mu'_{s,2}$ fixed at 1.1 mm$^{-1}$). This allows for the recreation of both the relationship between $\Delta\mu_{a,SD\phi}$ and $\Delta\mu_{a,DSI}$ observed in Subjects 1 and 3 *in vivo*, as well as the recovered $\Delta\mu_{a,M}$ magnitude difference between subjects for all $M$. Figure 8(d) shows results for a three-layer medium where only $L_1$ is varied, with $L_1 = 10$ mm (magenta) or $L_1 = 14$ mm (green). Like the two-layer medium results of Figure 8(c), the three-layer medium results of Figure 8(d) recreate both the relationship between $\Delta\mu_{a,SD\phi}$ and $\Delta\mu_{a,DSI}$ observed between Subjects 1 and 3 *in vivo*, as well as the recovered $\Delta\mu_{a,M}$ magnitude difference between subjects for all $M$ (but, in the three-layer medium case, for the same value of $\mu'_{s,1}/\mu'_{s,3}$). The effectively homogenous absolute optical properties of the *in vivo* data and the simulated scenarios are included in Table 1.

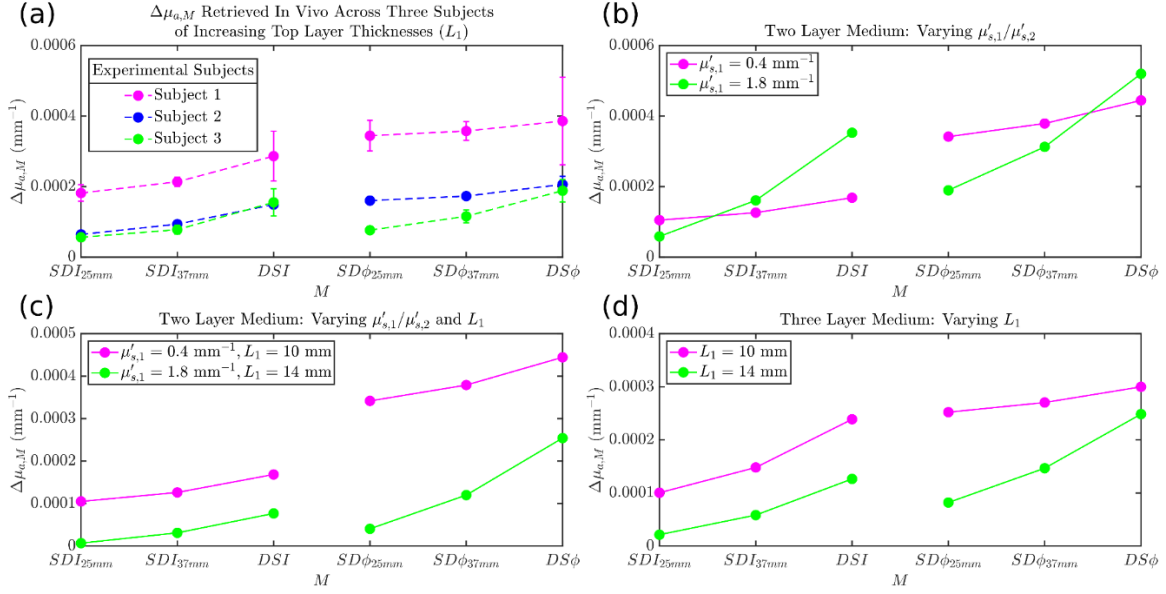

Figure 8. (a): $\Delta\mu_{a,M}$ retrieved *in vivo* from three different subjects (Subjects 1, 2, and 3), ordered in increasing thickness of the superficial layer (scalp+skull thickness). Error bars denote the standard error of the mean of each $M$. (b): $\Delta\mu_{a,M}$ calculated from two-layer simulations, one with a low scattering superficial layer ($\mu'_{s,1}$ = 0.4 mm$^{-1}$) and one with a highly scattering superficial layer ($\mu'_{s,1}$ = 1.8 mm$^{-1}$). Properties held constant: $\Delta\mu_{a,2}$=0.001 mm$^{-1}$, $\mu_{a,1}$ = 0.017 mm$^{-1}$, $\mu_{a,2}$ = 0.021 mm$^{-1}$, $\mu'_{s,2}$ = 1.1 mm$^{-1}$, $L_1$=10 mm. (c): $\Delta\mu_{a,M}$ calculated from two-layer simulations, one with a low scattering superficial layer ($\mu'_{s,1}$ = 0.4 mm$^{-1}$) and $L_1$=10 mm, and one with a highly scattering superficial layer ($\mu'_{s,1}$ = 1.8 mm$^{-1}$) and $L_1$=14 mm. Properties held constant: $\Delta\mu_{a,2}$=0.001 mm$^{-1}$, $\mu_{a,1}$ = 0.017 mm$^{-1}$, $\mu_{a,2}$ = 0.021 mm$^{-1}$, $\mu'_{s,2}$ = 1.1 mm$^{-1}$. (d): $\Delta\mu_{a,M}$ calculated from three-layer simulations, one with $L_1$=10 mm, and one with $L_1$=14 mm. Properties held constant: $\Delta\mu_{a,3}$=0.001 mm$^{-1}$, $\mu_{a,1}$ = 0.017 mm$^{-1}$, $\mu_{a,2}$ = 0.0027 mm$^{-1}$, $\mu_{a,3}$ = 0.021 mm$^{-1}$, $\mu'_{s,1}$ = 1.45 mm$^{-1}$, $\mu'_{s,2}$ = 0.01 mm$^{-1}$, $\mu'_{s,3}$ = 1.1 mm$^{-1}$.

Table 1. Effective homogenous absolute optical properties at 830 nm for Figure 8.

| | | Effective Homogeneous Absolute Optical Properties at 830 nm | | Superficial Layer Thickness *In vivo*: Scalp+Skull Range Simulation: $L_1$ |
|---|---|---|---|---|
| | | $\mu_a$ (mm$^{-1}$) | $\mu'_s$ (mm$^{-1}$) | Thickness (mm) |
| *In vivo* subjects as in Fig. 8(a) | Subject 1 | 0.0147(2) | 0.76(4) | 10.9(9) |
| | Subject 2 | 0.0112(7) | 0.91(3) | 11.6(4) |
| | Subject 3 | 0.0095(3) | 0.85(3) | 16(1) |
| Two-layer medium as in Fig. 8(b) | $\mu'_s$=0.4 mm$^{-1}$ | 0.0214 | 0.37 | 10 (magenta line) |
| | $\mu'_s$=1.8 mm$^{-1}$ | 0.0173 | 1.64 | 10 (green line) |
| Two-layer medium as in Fig. 8(c) | $\mu'_s$=0.4 mm$^{-1}$ | 0.0214 | 0.38 | 10 (magenta line) |
| | $\mu'_s$=1.8 mm$^{-1}$ | 0.0176 | 1.73 | 14 (green line) |
| Three-layer medium as in Fig. 8(d) | $\mu'_s$=1.45 mm$^{-1}$ | 0.0144 | 0.50 | 10 (magenta line) |
| | $\mu'_s$=1.45 mm$^{-1}$ | 0.0132 | 1.31 | 14 (green line) |

Errors in absolute optical properties are reported as the standard deviation taken spatially across all absolute optical properties in analyzed module. *In vivo* superficial layer thicknesses reported are the means found by considering a range of values of the ultrasound scaling factor to account for uncertainty in the speed of ultrasound in the scalp and skull layers. The error in the *in vivo* superficial layer thickness is the propagated standard deviation from the scalp and skull layers. All errors on the last significant digit are shown in parentheses.

## 4. Discussion

*4.1. Observed influences of layer properties in simulated functional activation*

The first observation made across all simulations is the consistent reduction of the recovered absorption change ($\Delta\mu_{a,M}$) as the superficial layer thickness ($L_1$) increases. While expected, the rate of this attenuation varies significantly by data-type. Regardless of the optical properties considered in the layered structures, across both $I$ and $\phi$ data-types and for all simulations, it is always true that $\Delta\mu_{a,\text{SD}_{25\,\text{mm}}} < \Delta\mu_{a,\text{SD}_{37\,\text{mm}}} < \Delta\mu_{a,\text{DS}}$. This observation demonstrates the enhanced depth sensitivity of DS data compared to SD data and further reinforces the conclusions of previously published DS work.[18,33] Additionally, it is observed, especially for three-layered media, that increasing $L_1$ can lead to differing orders of $\Delta\mu_{a,M}$ from greatest to smallest, as is additionally seen in the *in vivo* results. This can be due to the influence of the low scattering and low absorbing CSF-mimicking layer, and the way the structure of the probed media can influence the sensitivity regions of each $M$, therefore leading to varying $\Delta\mu_{a,M}$ values.[35]

## 4.1. Varying $\mu'_{s,1}/\mu'_{s,2}$ or $\mu'_{s,1}/\mu'_{s,3}$ and $\mu_{a,1}/\mu_{a,2}$ and $\mu_{a,1}/\mu_{a,3}$ by changing first layer vs. bottom layer properties

In exploring the impact of changing optical properties of each layer (for both the two- and three-layered medium) on the recovered $\Delta\mu_{a,M}$ values, we varied $\mu_{a,1}$ or $\mu'_{s,1}$, while leaving $\mu_{a,2}$ or $\mu_{a,3}$ and $\mu'_{s,2}$ or $\mu'_{s,3}$ constant. Additional sets of simulations were conducted where $\mu'_{s,1}/\mu'_{s,2}$ was varied through holding $\mu'_{s,1}$ constant and varying $\mu'_{s,2}$ for the two-layered medium, with additional sets conducted varying $\mu'_{s,1}/\mu'_{s,3}$ by holding $\mu'_{s,1}$ constant and varying $\mu'_{s,3}$ for the three-layered medium. The same procedure was followed for exploring variation in $\mu_a$ values. These additional simulation sets resulted in the same key takeaways as those included in Section 3.1, where we observed a decrease in $\Delta\mu_{a,M}$ for all $M$ as $L_1$ increases, a variation in $\Delta\mu_{a,M}$ magnitude order across $M$ as $L_1$ increases, and a higher influence on $\Delta\mu_{a,M}$ values and relationships through changing $\mu'_{s,1}/\mu'_{s,2}$ (or $\mu'_{s,1}/\mu'_{s,3}$) than changing $\mu_{a,1}/\mu_{a,2}$ (or $\mu_{a,1}/\mu_{a,3}$).

These results showed that the most influential optical properties in determining the relationship between the recovered $\Delta\mu_{a,M}$ associated with deep absorption changes is the ratio $\mu'_{s,1}/\mu'_{s,2}$ (or $\mu'_{s,1}/\mu'_{s,3}$), i.e. the relative scattering properties of the superficial and deep tissue layers. We also found that $\mu_{a,1}/\mu_{a,2}$ (or $\mu_{a,1}/\mu_{a,3}$) has a larger impact on the recovered $\Delta\mu_{a,M}$ values than the individual values of $\mu_{a,1}$ and $\mu_{a,2}$ (or $\mu_{a,3}$), although neither made a significant difference on the relationship of the recovered $\Delta\mu_{a,M}$ values to each other. This confirms that accurate forward modeling relies less on the absolute values of the optical properties of the various layers, and more heavily on correctly capturing the relative scattering properties of the various layers.

*4.2. Inclusion of CSF in three-layer simulations and CSF optical properties*

There has been much uncertainty throughout literature in regard to what optical properties best represent the CSF, and what ways of modeling elicit the most accurate depiction of the CSF in subarachnoid space. While pure, healthy CSF is clear and water-like, the complex arachnoid trabeculae and vasculature within the subarachnoid space introduce scattering.

Early work from Okada and Delpy modeled the CSF as a low-scattering layer ($\mu'_s$=0.001 mm$^{-1}$) with a thickness of 2 mm and $\mu_a$=0.002 mm$^{-1}$.[3,4] Alternatively, Custo *et al.* treated the CSF layer as a diffuse medium with $\mu'_s$=0.3 mm$^{-1}$.[23] Custo *et al.* argued that the use of this artificially higher scattering coefficient (up to the inverse of the typical line-of-sight distance) allows for the use of the diffusion approximation. These optical properties have been debated for potential overestimation (in the case of Custo's work) or underestimation (in the case of Okada and Delpy's work) of the true scattering of the CSF-filled subarachnoid space. Okada and Delpy later suggested that the presence of the trabeculae would justify the use of higher $\mu'_s$ values in the range of 0.16-0.32 mm$^{-1}$. More recently, Lewis and Fang questioned whether these values were overestimations, suggesting the use of $\mu'_s$=0.15 mm$^{-1}$ while maintaining physiologically relevant $\mu_a$ values (approximately $\mu_a$=0.0004 mm$^{-1}$ at 690 nm, and $\mu_a$=0.0026 mm$^{-1}$ at 830 nm).[22]

In response to this range of reported values, we conducted additional simulations varying the $\mu'_s$ used in the CSF-mimicking layer of our three-layer simulations, and found that using values anywhere in the range reported in the literature (0.001-0.32 mm$^{-1}$) elicits the same results as those presented in Sections 3.1, and allows for the reproduction of observed $\Delta\mu_{a,M}$ *in vivo* results. Our takeaway here is that the most important parameter for the inclusion of CSF is the ratio between the $\mu'_s$ of the scalp/skull or cerebral layer and the CSF-mimicking layer ($\mu'_{s,1}/\mu'_{s,2}$, $\mu'_{s,3}/\mu'_{s,2}$). As long as $\mu'_{s,1}/\mu'_{s,2} \gg 1$ and $\mu'_{s,3}/\mu'_{s,2} \gg 1$, the three-layer model that incorporates the CSF contribution to the optical signal is able to reproduce the *in vivo* results of $\Delta\mu_{a,M}$.

### 4.3. Effect of CSF thickness on in vivo trend reproduction

An open question regarding the inclusion of CSF in layered modeling is the thickness of the CSF layer itself, and the impact that this change can have on the recovered data. This is important due to the fact that the thickness of the CSF layer is significantly variable based on the region of the head being probed.[36,37] To address this issue, we ran simulations for physiological CSF thicknesses (2 or 3 mm) with the same optical properties discussed above. We found that the thickness of the CSF layer has a minor effect on the retrieved $\Delta\mu_{a,M}$ values, and does not affect the reproducibility of the observed *in vivo* extracerebral thickness-dependent trends. That being said, it is noticed that changing the thickness of the CSF layer has a much more notable effect on layered media with a thinner top layer (10 mm) than a thicker top layer (14 mm). This is reasonable since in layered media with a thinner top layer one would expect a higher likelihood of detecting photons that have traveled through the CSF layer itself.

*4.4. Effect of $\Delta\mu_a$ elicited theoretically on retrieved $\Delta\mu_{a,M}$*

While all of the simulations in this work included only a $\Delta\mu_a$ in the deepest layer of the layered structure (second layer for the two-layered medium, third layer for the three-layered medium), it is possible that functional protocols also elicit hemodynamic changes in superficial extracerebral tissue.[25–27,38,39] While this effect has been reported to be less prominent in visual stimuli tasks, we investigated absorption changes in the top layer ($\Delta\mu_{a,1}$) in combination with absorption changes in the deep cerebral layer ($\Delta\mu_{a,2}$ for two-layered medium, $\Delta\mu_{a,3}$ for three-layered medium). These results are shown in Figures 9 and 10. Figure 9 reports the effect on $\Delta\mu_{a,M}$ of $\Delta\mu_{a,1}$ and $\Delta\mu_{a,2}$ in a two-layer medium across different values of $L_1$. Here, we can see that as long as $\Delta\mu_{a,1}<\Delta\mu_{a,2}$, we retrieve the same qualitative relationships between $\Delta\mu_{a,M}$ for each $M$ as observed for $\Delta\mu_{a,1} = 0$. Figure 10 reports the effect on $\Delta\mu_{a,M}$ of $\Delta\mu_{a,1}$ and $\Delta\mu_{a,3}$ in a three-layer medium across different values of $L_1$. Here, we can observe that while the relationship between $\Delta\mu_{a,DSI}$ and $\Delta\mu_{a,SD\phi}$ is upheld as long as $\Delta\mu_{a,1}< \Delta\mu_{a,3}$ for thicker values of $L_1$, the order of the $\Delta\mu_{a,M}$s retrieved by the $I$ data-types as well as the $\phi$ data-types changes. The increase in $\Delta\mu_{a,M}$ obtained with SD$I_{25mm}$, SD$I_{37mm}$, and DS$I$, respectively, and with SD$\phi_{25mm}$, SD$\phi_{37mm}$, and DS$\phi$, respectively, is consistently observed in *in vivo* data, and is not reproduced by the three-layer medium simulations where $\Delta\mu_{a,1} \neq 0$. Therefore, we find that the case where only $\Delta\mu_{a,3}$ is relevant, or is accompanied by a significantly smaller $\Delta\mu_{a,1}$, is consistent with our *in vivo* results. This is also consistent with prior research, which found little to no detected systemic superficial hemodynamic change in visual stimulation protocols when compared to other functional protocols (cognitive, etc.).[38,40] However, since systemic superficial hemodynamic changes can be induced during some functional activation protocols or may result from concurrent systemic physiological changes, further investigation of *in vivo* functional data collected under more general conditions is an important future direction.

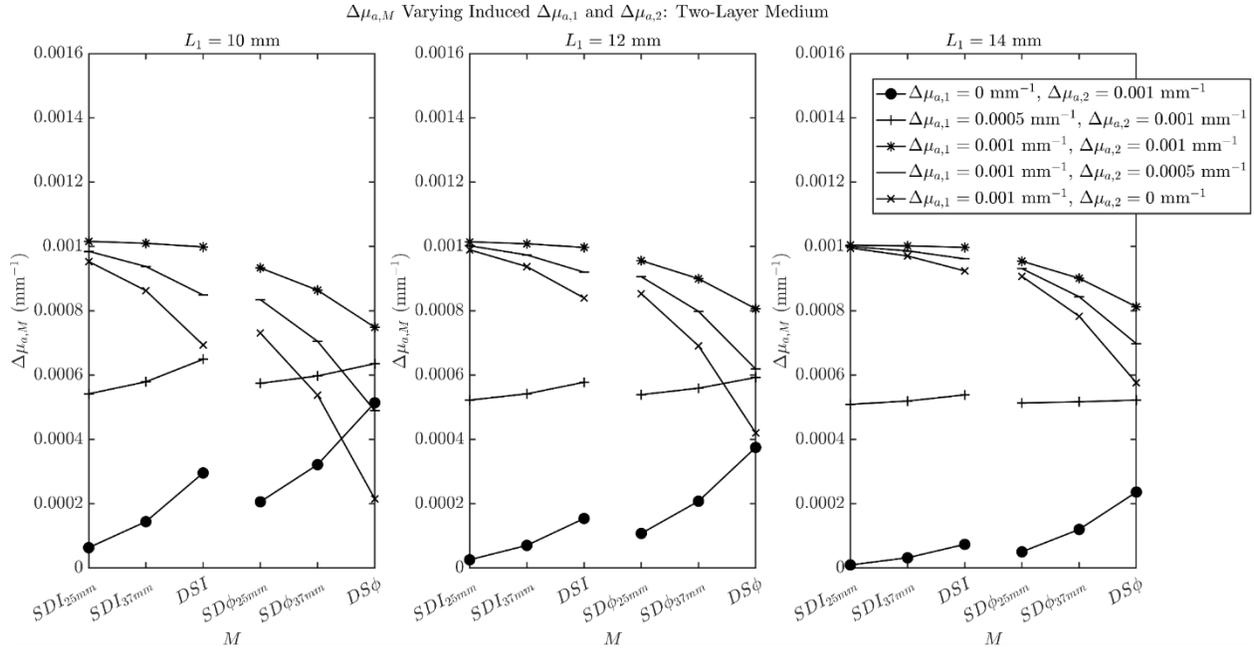

Figure 9. Varying $\Delta\mu_{a,1}$ and $\Delta\mu_{a,3}$ in differing amounts, in accordance with the shown legend, measuring the effect of the retrieved $\Delta\mu_{a,M}$ for three different top layer thicknesses: $L_1$ = [10, 12, 14] mm. Properties held constant: $\mu_{a,1}$ = 0.017 mm$^{-1}$, $\mu_{a,2}$ = 0.021 mm$^{-1}$, $\mu'_{s,1}$ = 1.45 mm$^{-1}$, $\mu'_{s,2}$ = 1.1 mm$^{-1}$.

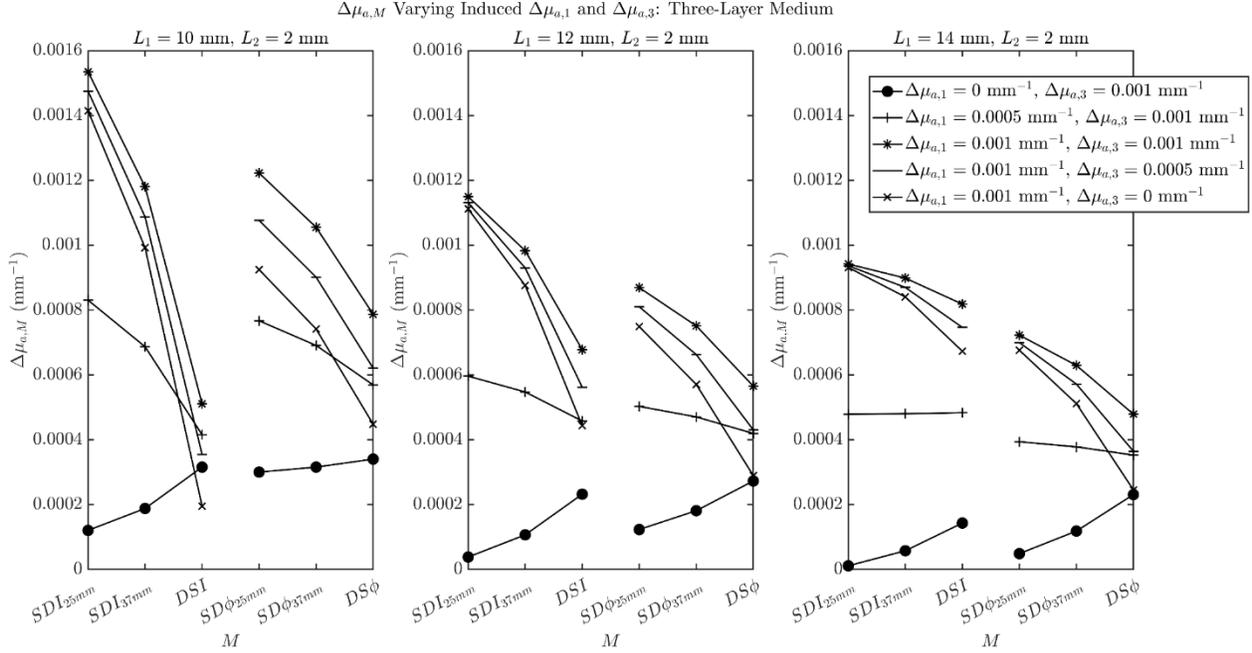

Figure 10. Varying $\Delta\mu_{a,1}$ and $\Delta\mu_{a,3}$ in differing amounts, in accordance with the shown legend, measuring the effect of the retrieved $\Delta\mu_{a,M}$ for three different top layer thicknesses: $L_1$ = [10, 12, 14] mm. Properties held constant: $\mu_{a,1}$ = 0.017 mm$^{-1}$, $\mu_{a,2}$ = 0.0027 mm$^{-1}$, $\mu_{a,2}$ = 0.021 mm$^{-1}$, $\mu'_{s,1}$ = 1.45 mm$^{-1}$, $\mu'_{s,2}$ = 0.01 mm$^{-1}$, $\mu'_{s,2}$ = 1.1 mm$^{-1}$.

*4.5. Varying $\mu'_{s,1}/\mu'_{s,2}$ in a two-layer medium vs. varying $L_1$ in a three-layer medium to reproduce in vivo results*

In exploring the *in vivo* $\Delta\mu_{a,M}$ values measured in three subjects during visual stimulation, we identified two medium configurations that could reproduce most features of the *in vivo* results. The first one is a two-layer medium where both $L_1$ and $\mu'_{s,1}/\mu'_{s,2}$ had to be significantly varied across subjects, and the second one is a three-layer medium where only $L_1$ needed to be varied across subjects. The two-layer medium required $L_1$ to be thinner for Subject 1 than Subject 3 (which is in agreement with the ultrasound imaging measurements of superficial extracerebral tissue thickness for these subjects, see Table 1), whereas $\mu'_{s,1}/\mu'_{s,2}$ had to be <1 for Subject 1 and >1 for Subject 3. Such difference in the relative scattering properties of extracerebral and cerebral tissues in the two subjects is biologically questionable. While scalp and skull optical properties can vary on an individual basis, this variation has been reported to be significantly smaller[3,4] than required for our two-layer simulations to reproduce the *in vivo* results. Additionally, when comparing the effectively homogeneous absolute optical properties for the three subjects (shown in Table 1), the effectively homogenous $\mu_s'$ values retrieved are significantly more similar to one another (standard deviation of 0.1045 between Subject 1 and Subject 3) than that of the effectively homogenous $\mu_s'$ values retrieved for the two-layer simulated case that recreates the $\Delta\mu_{a,M}$ relationship *in vivo* (standard deviation of 0.8968). While the effectively homogenous $\mu_s'$ values retrieved for the three-layer simulated case that recreates the $\Delta\mu_{a,M}$ relationship *in vivo* still do have more variation between each other than that of the *in vivo* case, there is significantly less variation than that of the two-layer case (standard deviation of 0.5781). Therefore, we posit that the three-layer medium, where the second layer represents the CSF, is a more accurate model than the two-layer medium to reproduce the *in vivo* results, as it only requires the adjustment of $L_1$ across subject. This finding leads to our recommendation to use a three-layered tissue model to analyze DS FD-NIRS data to recover $\Delta\mu_a$ values that are most indicative of cerebral hemodynamics, which is our main overarching goal.

*4.6. Limitations of the layered model*

While this work is based on a layered tissue model to reproduce DS FD-NIRS data-types collected *in vivo*, and to identify ways to leverage such data-types to obtain measurements of $\Delta\mu_a$ that are most representative of cerebral hemodynamics, we realize that the layered model is still a simplification of the tissue anatomical complexity. Additionally, the layered structure does not allow for the inclusion of laterally inhomogeneous $\Delta\mu_a$ perturbations, which is the biological reality of the vascular network structure and of localized functional activation.[40] As this work is solely focused on functional activation from visual stimuli that map across the primary visual cortex and elicit spatially extended hemodynamic responses, we believe that the use of a layered model in this case is appropriate, certainly as an advancement over a homogeneous model.[41] That being said, future work will focus on incorporating lateral inhomogeneities to more closely represent the tissue structure and functional responses, including the incorporation of sulci when examining the contributions of CSF.

## 5. Conclusion

This work has leveraged the large information content, in terms of range of sensitivities to the tissue optical properties and their spatial distribution, of DS FD-NIRS data, which include intensity and phase data in single-distance (SD) configuration (either 25 or 37 mm source-detector separation), and in dual-slope (DS) configuration (combination of two SD data at 25 mm and two SD data at 37 mm). A key question investigated in this study is whether DS FD-NIRS data generated with Monte Carlo simulations for two- and three-layered media are able to reproduce experimental data collected *in vivo* over the occipital lobe of human subjects during a visual stimulation protocol. The main result of this work is that a three-layered medium, with a second layer that is less absorbing and less scattering than the other two layers, and with a top layer thickness that represents the combined scalp and skull thickness, is able to reproduce the main qualitative features of *in vivo* data.

This research indicates that a three-layer model can be used to analyze functional DS FD-NIRS data (and, more generally, any fNIRS data) as a significant improvement upon homogenous models to obtain more accurate measurements of cerebral hemodynamic without a need for large data sets for tomographic reconstructions. In principle, the three-layer model requires the determination of 14 parameters (absorption and scattering properties of each layer at two wavelengths, thickness of the top two layers), but the results of this work provide some guidance for the selection of these parameters. The values of $\mu_{a,1}$ and $\mu_{a,3}$ can be taken within the range of 0.005-0.025 mm$^{-1}$, with both their values and wavelength dependence being linked to the concentration and oxygen saturation of hemoglobin in these tissue layers. The values of $\mu'_{s,1}$ and $\mu'_{s,3}$ can be taken within the range of 1.0-1.5 mm$^{-1}$, which have been shown to be robust values for reproducing the *in vivo* results. Both absorption and reduced scattering coefficients of the CSF layer ($\mu_{a,2}$ and $\mu'_{s,2}$) should be much less than the corresponding coefficients in the first and third layers, with guidance from the values used in this work ($\mu_{a,2} \approx 0.0027$ mm$^{-1}$, $\mu'_{s,2} \approx 0.01$ mm$^{-1}$). Of course, time-resolved and spatially-resolved NIRS data at baseline can provide valuable information toward the assignment of optical properties of the various tissue layers. In terms of layer thicknesses, $L_1$ may be estimated by measuring the depth of the skull base with ultrasound imaging, and $L_2$ should be $\ll L_1$ (in this work, $L_2 \approx 2$ mm). By populating a forward model with these parameters, one may calculate specific generalized partial path lengths within a tissue layer (i.e. the partial derivatives of any NIRS data-type with respect to the absorption of any tissue layer) to assign the contributions from absorption changes in each layer to changes in measured NIRS data. This approach is a paradigm for simultaneous recovery of $\Delta\mu_{a,1}$ (scalp hemodynamics) and $\Delta\mu_{a,3}$ (cerebral hemodynamics), which addresses the need to account for superficial hemodynamics contributions to fNIRS data and perform more accurate measurements of cerebral hemodynamics.

**Code and Data Availability**

Data and code will be made available at the following repository: https://github.com/DOIT-Lab/DOIT-Public/tree/master/ThreeLayer_BiophotonicsDiscovery.

**Acknowledgements**

The authors would like to acknowledge support from the National Institutes of Health R01-EB029414. GB is funded by NIH K99-HL181290 and would also like to acknowledge support from NIH K12-GM133314. The content is solely the authors' responsibility and does not necessarily represent the official views of the awarding institutions.

**Disclosures**

The authors disclose no conflicts of interest.